\documentclass[twocolumn,showpacs,aps,pre]{revtex4}
\usepackage{graphicx}
\usepackage{dcolumn}
\usepackage{bm}

\begin{document}

\title{Friction between Polymer Brushes}
\author{J. B. Sokoloff}
\address{Physics Department and Center for Interdisciplinary Research 
in Complex Systems,
 Northeastern University, Boston, MA 02115.}

\date{\today }

\begin{abstract}
By solving the equilibrium equations for a polymer in a neutral polymer 
brush, the degree of 
interpenetration of two polymer brushes in contact and near contact is 
calculated. These results are used to calculate values of the force of 
static friction in agreement with recent friction measurements 
for polymer brush lubricated surfaces. It is shown that at sufficiently 
light loads polymer brush coated surfaces can slide, with the load 
supported entirely by osmotic pressure, at a sufficiently 
large spatial separation so as to avoid entanglement, and hence static 
friction. 
\end{abstract}


\maketitle

\section{Introduction}
A polymer brush consists of a surface with a fairly concentrated coating 
of polymer chains, each one of which has one end tightly bound 
to the surface. They serve as extremely effective lubricants, producing 
friction coefficients  as low as 0.001 or less!  Polymer brushes are a 
promising way to reduce friction to extremely low values.   They have 
the disadvantage, however, that they must be immersed in a liquid solvent 
in order to function as a lubricant.    
Since joints in the human body are known to be immersed in a fluid, 
known as the synovial fluid, it is likely that they are lubricated by 
polymer brushes, where  polymers known as hyaloronan or lubricin are 
attached to the cartilage which coats the bones making up the joints\cite{klein}.   
The density profile of a polymer brush (i.e., the density of monomers 
as a function of distance from the surface to which the polymers are 
attached) is well established\cite{milner,witten}. What is not understood is how the 
interaction of polymer brush coated surfaces in contact with each 
other is able to account for the observed friction.   
For example, molecular dynamics studies generally do not predict static 
friction\cite {grest}, whereas surface force apparatus measurements 
due to Tadmor, et. al.\cite {tadmor}, find that there is static friction.

Pincus, Witten, Milner, etc. \cite{milner,witten},
present the following picture: over all except the very outer edge of a 
brush, the monomer density at a distance z from the solid surface is
proportional to $h_{max}^2-z^2$, but there is a low density tail for $z>h_{max}$, 
where $h_{max}$ is the height of an uncompressed polymer brush. 
When two brushes are put in contact the brushes are compressed and the 
resulting elastic force (which is really of entropic origin) and the osmotic pressure 
support the load. The osmotic  
pressure is not of the usual textbook example in which a solute is trapped
in an impenetrable membrane because here what plays the role of the
membrane are the forces that hold the monomers together in each polymer    
chain. Since each polymer is anchored to a one of the two surfaces, it   
cannot escape from the region between the two surfaces. Solvent is pulled 
into the region between the plates, diluting the monomer density (which 
plays the role of a solute density) resulting in osmotic pressure. 

 The experiments of Tadmor, et. al. \cite {tadmor} exhibit static  
friction which increases logarithmically with the time that the polymer 
coated surfaces are in contact. The logarithmic increase of
static friction with resting time is qualitatively similar to a similar   
effect for solid-solid contact, even though the mechanisms for it are much
different\cite {baumberger}.

     Kinetic friction is discussed in a paper by Joanny\cite {joanny}, 
assuming that it occurs by reptation of polymers from one brush
that get entangled in the second brush. He finds only a viscous friction, 
but if the polymers near the interface get entangled enough, there must 
be something that resembles static friction if the times scale associated 
with the reptation motion is very
long compared to the time for the surfaces to move some very small   
distance in creep motion as they attempt to slide. Even for solid-solid friction, 
there is always creep motion. Creep, however, 
for the polymer brushes happens over a much shorter time scale than for   
solid-solid contact. 

In order to calculate the static friction it is necessary to determine the 
number of monomers belonging to polymers of one brush that penetrate 
into the 
second brush. This can be accomplished using an extension of the approach used in 
Ref. \cite {milner}. Consider two polymer brushes which are pressed together. 
The mean field potential for the hard core interaction of a monomer 
belonging to a polymer under consideration with other monomers is given by 
$$w[\phi (z)+\phi (D-z)], \eqno (1)$$
where D is the separation between the surfaces, to which the two polymer 
brushes that we are considering are attached, z is the distance from one 
of the two surfaces, which we will consider to be the lower surface,  
$\phi (z)$ is the monomer density of one of the polymer brushes and 
w is the standard monomer excluded volume parameter\cite {deGennes}. 
In equilibrium, $\phi$ has the standard parabolic density\cite {milner} 
$$w\phi (z)=B(h_{max}^2-z^2), \eqno (2)$$
for $z<h$ and zero for $z>h$, where h is the actual height of the brush and 
where $h_{max}$ is its maximum height, which is the equilibrium height when 
the brush is not in contact with a second brush and $B=\pi^2 k_B T/(8a^2 N^2)$, 
the form that B must have in order for a polymer attached on one end to 
the surface at z=0 to extend to an arbitrary height $z=\rho<h$. 
In Ref. \cite{milner}, it is shown that 
$h_{max}=(12/\pi^2)^{1/3} (a^2\sigma w/k_B T)^{1/3} N$, where the factors of 
$k_B T$ and a, which were set equal to one in Ref. \cite{milner}, have 
been inserted here. Here, a is the intermonomer distance, $k_B$ is Boltzmann's 
constant, T is the absolute temperature and $\sigma$ is the number of 
polymers belonging to the polymer brush attached to the surface per unit 
area. When the brushes are 
in contact and the two brushes are pushed together under a load, $h<h_{max}$, 
which means that the brushs get "flattened out" so that $\phi (z)$ no longer 
goes to zero at z=h.  
For infinite N, all of the monomers belonging to the polymers belonging 
to the lower brush (i.e., the polymers attached to the lower 
surface) remain within that brush (i.e., in the region for which $z<h$), 
and the monomers belonging to the polymers of the upper brush (i.e., 
the polymers attached to the upper surface) remain in that brush 
or in the region defined by $h<z<D$, where D=2h). For finite values 
of N, however, there exist thermal fluctuations in which polymers from 
the lower brush penetrate into the upper brush and polymers belonging 
to the upper brush penetrate into the lower one. In order to calculate  
the degree of penetration, let us consider the free energy needed to 
add a polymer to one of the brushes \cite {milner}, 
$$F(k)-F(k-1)=\int_0^N dn [(1/2)(k_B T/a^2)|{d{\bf r_n}\over dn}|^2+$$
$$w(\phi (z_n)+\phi (D-z_n))], \eqno (3)$$
where F(k) is the free energy for a two brush system containing k 
polymers and $r_n$ is the location of the $n^{th}$ monomer in 
the polymer that was added to the lower brush\cite {milner}. We can 
find the a local minimum value of the single polymer free energy in Eq. (3) by 
setting the variation of this expression with respect to ${\bf r_n}$ 
equal to zero, which gives
$$(k_B T/a^2){d^2 z_n\over dn^2}=w[{d\phi (z_n)\over dz_n}+{d\phi (D-z)\over dz_n}]  
\eqno (4)$$
and $dx_n/dn=dy_n/dn$ are constant. Since we are interested in small 
fluctuations from the infinite N configuration of the brushes (for 
which the brushes do not interpenetrate), to lowest order in 1/N, we may use 
the parabolic monomer density of Eq. (2), which was found in the infinite N 
limit. Solutions to this problem for which the polymer under consideration, 
which is attached to the lowest surface, penetrates into the region for 
which $z>h$ may be found by finding such solutions to Eq. (4). For $z<h$, Eq. (4) 
is 
$${d^2 z\over dn^2}=w{d\phi (z)\over dz}=-2Bz \eqno (5a)$$ 
and for $h<z<D$, it is
$${d^2 z\over dn^2}=w{d\phi (D-z)\over dz}=-2B(z-D). \eqno (5b)$$
The solution of Eq. (5a) is
$$z=ccos (\omega n) \eqno (6a)$$
and the solution to Eq (5b) is 
$$z=D+acos (\omega n)+bsin(\omega n) \eqno (6b)$$
where a, b, and c are constants and the "frequency" $\omega=(2Ba^2/k_B T)^{1/2}=\pi/(2N)$. 
The choice of the cosine solution in Eq. (6a) was made because for a 
polymer attached to the lower surface $z_N=0$ and the expression for 
$\omega$ under Eq. (6b) will result in this condition being satisfied if we \
choose the cosine solution. If we assume that for $n>k_1$, the value of n for the 
monomer which first penetrates into the second brush, we must demand that the 
values of z given by Eq.'s (6a) and (6b) and their derivatives with respect to n 
must be equal at $n=k_1$, since these are second order differential equations. 
This condition gives 
$$a=-h[cos(\omega k_1)-sin(\omega k_1)tan(\omega k_1)], \eqno (7a)$$
$$b=-2hsin(\omega k_1), \eqno (7b)$$ 
$$c={h\over cos(\omega k_1)}. \eqno (7c)$$
Since we expect for large N and low temperature that $k_1<<N$, we find from Eq. (6b), 
using Eq. (7) that 
$$z_0-h\approx (3/2)h(\omega k_1)^2 \eqno (8a)$$
and 
$${dz_n\over dn}|_{n=0}\approx -2h\omega^2 k_1. \eqno (8b)$$
Following the discussion in Ref. \cite {milner}, we find that the fluctuation in 
the free energy of a polymer belonging to the lower brush when $k_1$ of its 
monomers penetrate into the upper brush is equal to the amount of work by the tension 
that would have to be applied to the n=0 monomer to pull it out of the lower brush 
by an amount $y=z_0-h$, which is given by
$$\Delta S=(k_B T/a^2)\int_h^{h+y}{dz_n\over dn}|_{n=0}dz_0. \eqno (9)$$
If we change variable of integration to $k_1$ using Eq.'s (8), we obtain 
$$\Delta S=2 (k_B T/a^2)\omega^4 h^2 k_1^3. \eqno (10)$$
Using Eq. (10), we find that the probability of $k_1$ monomers of a polymer 
from the lower brush penetrating into the upper brush is given by the 
Boltzmann factor
$$e^{-\Delta S/K_B T}=exp[-2\omega^4(h/a)^2 k_1^3]. \eqno (11)$$
Thus, we conclude that the mean length of a segment of a polymer belonging 
to one brush that gets entangled in the second brush is $<k_1>$ found from Eq. (14), 
which is of order $[(1/2)(a/h_{max})^2\omega^{-4}]^{1/3}$ or larger (for $h<h_{max}$, 
which is of order $N^{2/3}$, and hence these segments are sufficiently long for large N to form 
"blobs" which are close packed in the brush that it has penetrated. This is 
expected to occur because the polymers belonging to the two brushes are in 
the semi-dilute regime \cite {deGennes}.  This must be the case because the Flory radius 
$R_F$ of a polymer belonging to a polymer brush\cite {milner}is always greater than the 
mean spacing of the polymers, and hence the 
polymers belonging to the brushes cannot individually curl up into non-overlapping 
"blobs" of radius $R_F$, as they would in the dilute limit. The expression for the 
monomer density in a polymer brush $\phi (z)$ given by Eq. (2), can be written as 
$\phi (z)=(N\sigma/h_{max})[1-(z/h_{max})^2]$. The mean value of the monomer density in the 
region of the brush into which polymers belonging to the brush in which it is in contact 
have penetrated is very close to $c=\phi (z=h)$. Then, from Ref. \cite {deGennes}, 
the mesh size in the brush is approximately $\xi=a(ca^3)^{-3/4}=88a=2.64\times 10^{-6}cm$. 
Then the static friction force f per polymer 
that penetrates is the force needed to pull a "blob" through the mesh, which is 
$$f={k_B T\over \xi} \approx 1.52\times 10^{-8} dyn, \eqno (12)$$
which gives a force per unit area of $f\sigma\approx 2.15\times 10^3 N/m^2$.

Let us now consider the exciting possibility that at sufficiently light load the bulk of the two 
polymer brushes might not be in contact, meaning that $D>2h$, and that the load is 
entirely supported by osmotic pressure due to those polymers that extend out of each brush into a thin 
interface region of thickness $D-2h$. (Because of the osmotic pressure in the 
interface region, h will be compressed below $h_{max}$.) It will be shown that this interface region can be sufficiently thick so that the polymers from the two brushes do not entangle to any significant degree, resulting in negligibly small static friction.  Then, the end of the polymer $z_0$ is located 
in the interface region and for the part of the polymer in the interface region ($0<n<k_1$), 
$dz_n/dn=-v_1$, a constant, and hence $z_n=z_0-v_1 n.$ For $N>k_1$, $z_n$ is still given by 
Eq. (6a) and (7c). Again requiring continuity of $z_n$ and $dz_n/dn$ at $n=k_1$ and assuming that 
$\omega k_1<<1$, we obtain $v_1=h\omega k_1$, and hence,
$\Delta S=(k_B T/a^2)\int_0^{k_1}v_1 dz_0=(2/3)(k_B T/a^2)h^2\omega^4 k_1^3$ and hence the probability 
that $k_1$ monomers of the chain penetrate into the interface region is given by
$$P=exp[-(2/3)(h\omega^2/a)^2 k_1^3] \eqno (13)$$
Hence $<k_1>\approx (3/2)^{1/3} (a/h\omega^2)^{2/3}=1.12N^{2/3}$ for the values for $\sigma$ 
and N from Ref.\cite{tadmor} used here and taking $a\approx 3\times 10^{-8}cm$. In order for 
there to be no significant penetration of polymers belonging to one brush into the second, 
we require that $D-2h>z_0-h\approx h\omega^2 <k_1>^2$. Substituting for $<k_1>$ from the 
expression under Eq. (13) for it, using the parameters from Ref. \cite {tadmor}, we 
obtain $z_0-h\approx 47A^o$. We can estimate the monomer density in the interface 
region from $c\approx [\sigma/(D-2h)]<k_1>\approx [\sigma/(z_0-h)]<k_1>\approx 0.4\times 10^{21}cm^{-3}$. 
In the interface region for the parameters used here, we are in a dilute regime, in which 
the osmotic pressure is given by\cite {deGennes} $(1/2)k_B Ta^3 c^2$ for the above value of 
c is equal to $0.864\times 10^4 N/m^2$ at room temperature. Thus, at loads smaller than 
$0.864\times 10^4 N/m^2$ the polymer brush coated surfaces would be separated by a 
sufficiently thick interface region that there would be negligible entanglement 
of polymers belonging to the two brushes, and hence there would be negligible static 
friction. Charged polymer brushes have counterions which can provide an additional 
source of osmotic pressure, in addition to the net charge of the lubricated surfaces to 
support part of the load. This opens up the possibility that the load might be 
completely supported by osmotic pressure and electrical charge for charged polymer brushes. 

The normal force per polymer that can be supported by two 
brushes, each with a parabolic monomer distribution is the derivative of 
the free energy per polymer of one of the brushes given in Eq. (33) of Ref. 
\cite {milner}, which can be written as 
$$F=({N^2\sigma w\over h_{max}^2}){[0.9+1.5\Delta^3] \over N}, \eqno (14)$$
where $\Delta=(h_{max}-h)/h_{max}$. The resulting normal force $f_{N}$  
per unit area, which is equal to minus the derivative of F with respect 
to h multiplied by $\sigma$, is given by
$$f_N\sigma=(9/2)({N^2\sigma w\over h_{max}^2})\Delta^2.\eqno (15)$$
In the experiments of Ref. \cite {tadmor}, the surface coverage density 
of the polymers in the brushes is $(84 A^o)^{-2}=1.42\times 10^{12}cm^{-2}$. 
we determine that $h_{max}=451\times 10^{-8}cm$ from which we find that  
$f_N\sigma=0.811\times 10^5\Delta^2 N/m^2$, which is of the same order of 
magnitude as the load in Ref.\cite{tadmor}. Combining this and Eq. (12) 
and the discussion above it, it is found that the coefficient of static 
friction is greater than 0.0265, in agreement with Ref. \cite {tadmor}. If 
the surfaces to which the polymers are attached are perfectly smooth (e.g., 
for mica), $f\propto f_N^{3/8}$. As discussed earlier, we expect there to 
be creep for polymer brushes which occurs over a much shorter time scale 
than for typical lubricated surfaces. Let us consider what would happen 
if we apply a shear force below the maximum static friction to the two 
lubricated solids. On the average, an entangled polymer will work itself 
free in a time $\tau$, the reptation relaxation time\cite {deGennes}, 
resulting in the center of masses of the two surfaces sliding a distance 
$\Delta x/N_e$ relative to each other, where $\Delta x$ is of the order of 
the spacing between attached polymers in the brush and $N_e$ is the number 
of entangled polymers. Dividing this by $\tau$ and multiplying by $N_e$, 
we get a mean creep velocity of $\Delta x/(\tau)$. For the polymers 
of Ref. \cite{tadmor}, each of which contains about 1300 monomers, 
$\tau$ is found\cite {deGennes} to be about 0.02 s. As $\Delta x$ in 
Ref. \cite {tadmor} is about $84A^o$, we obtain a creep velocity 
of $4.2\times 10^{-5}cm/s$, so that the surfaces will slide by about 
1.5 mm in an hour.

The kinetic friction in the slow speed sliding limit, which must be 
comparable to the static friction can be found from the Tomlinson 
model \cite {tomlinson}, as follows: We may crudely treat an entangled 
polymer as a spring, one end of which is moving in a potential possessing 
several minima, taken to represent the potential due to polymers in the brush in 
which it is entangled. The opposite end of the spring 
is assumed to move with a sliding velocity v. Then the total potential of this polymer is 
given by
$$V=(1/2)\alpha (x-vt)+V(x), \eqno (16)$$
where $\alpha$ is the force constant of the spring, $V(x)$ is a  
potential which possesses several minima (which represents the energy needed to 
pull a polymer out of a brush in which it is entangled, and t is the time). If 
$\alpha<{d^2V(x)\over dx^2}|_{max}$, this potential possesses multiple minima\cite{tomlinson}. 
Here ${d^2V(x)\over dx^2}|_{max}$ is the maximum 
force constant of the bottom of the entanglement potential well. Following 
the treatment of a semi-dilute solution of polymers in Ref. \cite {deGennes}, 
the entanglement potential represents the energy that must be expended to pull 
a "blob" through the mesh created by the other polymers among which the polymer 
under consideration is entangled. In order to pull the polymer through the mesh, 
each "blob" must be stretched so that it will become small enough to fit through. 
It follows that the force constant of the mesh potential must be equal to 
the force constant for a single "blob." Since the chain as a whole consists of 
a bunch of "blobs" connected in series the effective force constant of the chain 
must be equal to the force constant of a single "blob" divided by the number 
of "blobs" of the polymer chain under consideration which are entangled in the 
other brush. Then clearly the above condition for multistability is always satisfied. 
As the chain 
is pulled through the potential a potential minimum containing the end of the chain 
periodically becomes unstable, allowing the end of the chain to drop 
into a potential well of lower energy. When this occurs, the resulting drop 
in potential energy gets converted into kinetic energy, which is assumed 
to get rapidly dissipated into excitations of the system. Setting the 
rate at which such energy dissipation occurs equal to the product of the 
kinetic friction and v, we determine that there must be kinetic 
friction in the low v limit. 

We conclude that entanglement of polymers belonging to one brush in 
the second brush is able to account for the static friction observed in 
Ref. \cite {tadmor} for two polymer brushes in contact. Although it was estimated 
that there is creep of about 1.5 mm per hour, on short time scales there 
does appear to be true static friction. Osmotic pressure alone 
was found to be able to support a load of the order of $10^{4}Pa$ while keeping 
the polymer brushes sufficiently far apart to prevent the entanglement that 
results in static friction.

\section*{Acknowledgment}
I wish to thank the Department of Energy (Grant DE-FG02-96ER45585).

\end{document}